\documentclass{llncs}
\usepackage{times}
\usepackage[T1]{fontenc}
 \usepackage{cite}

\usepackage{amssymb,amsmath}
\usepackage{wrapfig}
\usepackage{graphicx}
\usepackage{stfloats}
\usepackage{subfig}

 \newcommand{\outtitle}{Scalable Analysis for Large Social Networks: \\the data-aware mean-field approach}
 
 \newcommand{\trref}{the appendix} 
 \newcommand{\comment}[1]{#1} 

\makeatletter
\renewcommand{\paragraph}{%
  \@startsection{paragraph}{4}%
  {\z@}{1.0ex \@plus 1ex \@minus .2ex}{-1em}%
  {\normalfont\normalsize\itshape}%
}
\makeatother

\title{\outtitle{}}
\author{}
\institute{}
\author{%
Julie M. Birkholz\inst{1,3}\and Rena Bakhshi\inst{2,3}\and Ravindra Harige\inst{2,3}\and\\ Maarten van Steen\inst{2,3}\and  Peter Groenewegen\inst{1,3}}
\institute{Organization Sciences Department,
\email{{\textbraceleft j.m.birkholz, p.groenewegen\textbraceright}@vu.nl}
\and
Computer Science Department, \\ \email{rbakhshi@few.vu.nl, ravindra.harige@student.vu.nl, steen@cs.vu.nl}
\and
Network Institute, VU University Amsterdam, The Netherlands
}

\begin{document}

\maketitle
\pagestyle{plain}
\pagenumbering{arabic}

\begin{abstract}
Studies on social networks have proved that endogenous and exogenous factors influence dynamics. Two streams of modeling exist on explaining 
the dynamics of social networks: 1) models predicting links through network properties, and 2) models considering the effects of social 
attributes. In this interdisciplinary study we work to overcome a number of computational limitations within these current models. 
We employ a \emph{mean-field model} which allows for the construction of a population-specific model informed from empirical research for predicting links from both network and social properties in large social networks.. The model is tested on a population of conference 
coauthorship behavior, considering a number of parameters from available Web data. We address how large social networks can be modeled 
preserving both network and social parameters.  We prove that the mean-field model, using a data-aware approach, allows us to overcome 
computational burdens and thus scalability issues in modeling large social networks in terms of both network and social parameters. Additionally, 
we confirm that large social networks evolve through both network and social-selection decisions; asserting that the dynamics of networks cannot 
singly be studied from a single perspective but must consider effects of social parameters.
\end{abstract}


\section{Introduction}

\label{sec:intro}
Dynamics of social networks are receiving increasing attention in multiple research domains \cite{Ahuja2012,snijders:2010,barbasi02}. Theoretical developments posit that dynamics are influenced by network \cite{barabasialbert1999} and social processes \cite{snijders:2010}; with recent theory suggesting that the two co-evolve \cite{Ahuja2012}. Methods to explore dynamics of networks traditionally implement evolving graph models, using inferential statistics to assert the likelihoods of the creation, maintenance or dissolution of edges. Two distinct classes of modeling exist: 1) exclusively modeling the effect of network structures on dynamics \cite{dln,newman06}, and 2) modeling effects of social parameters and network effects for small networks ($\sim$ 1000 nodes) \cite{snijders:2010}. Both types of models prove that network processes affect the dynamics of networks. Network models have been able to accurately predict a small percentage of edges, suggesting that dynamics may also be fed by other processes. Social-parameter models have proved social attributes, in combination with network structures, play a role in network dynamics.

Despite this growing knowledge from both model classes, these models have limitations. The main limitation relates to using an evolving graph model which calculates statistical probabilities of individual nodes. This approach generally leads to a super-linear growth in computational load as the network size increases, partly caused by the quadratic growth in the number of links that need to be considered. Both models attempt to overcome this through different means. One is limited to either testing the effect of a few parameters on a large network, or a number of parameters on small networks. Consequently, neither provide a terrain to empirically confirm the effect of both network and social parameters in large social networks. 

In order to better understand the dynamics of large social networks, a different computational approach must be taken to overcome the issue of scalability in present models. 
In this paper we review the two existing model classes used to investigate dynamic social networks, and present a model for overcoming a number of acknowledged limitations. Using a mean-field model approach we are able to overcome scalability issues in previous models through aggregation of individual nodes. Parameters are developed using a data-aware approach which combines empirical research from Social Science and standard inferential statistics to develop a population-specific model for exploring the dynamics of collaboration in science. 

We consider the question whether mean-field modelling allows us to describe the behavior of a social system, considering a number of network 
and social parameters. In this first application of the mean-field model to large social networks, we aim to explain the effect of a set of parameters governing networking patterns of collaboration in Dutch Computer Science (CS).
Four parameters are considered in this research: institutional affiliation, scientific age, cosmopolitanism of knowledge production, and visibility of the scientists. 
We prove that mean-field models expand the empirical testing ground of dynamic network models through increased scalability.  
This allows us to better understand dynamics of large social networks, covering space that has not been 
investigated in the past using a mean-field approach.

The paper is set up as follows. In Section~\ref{sec:models} we review the state of social network models, specifically highlighting the limitations 
of present models. In Section~\ref{sec:model} we explain the mean-field model, discussing in detail the computational advantages of the model as 
well as the steps taken to implement a data-aware approach for improved specifications. In Section~\ref{sec:Application}, we test the model on 
the coauthorship networks of papers from the conference proceedings for Dutch computer scientists, collected from the DBLP data set for 2006 -- 2010. 
Finally, we conclude with the results and implications for scalable, data-aware modeling solutions for explaining dynamics of social networks.  
\section{Network Models}
\label{sec:models}
The evolution of a network is driven by the addition, maintenance, and dissolution of interactions (edges) between 
nodes over time. Evolving graph models are the most commonly implemented models to explain the dynamics of networks~\cite{newman2004cna,Barabasi2002590,grossman:03}. 
These models assume that nodes are added one-by-one to the network, in discrete time. They infer the probability of a link emerging given a 
node-transition rate using a Markovian model of simulation. Within this model type two distinct approaches exist 
investigating social network dynamics: 1) global network-structure link-prediction models, and 2) social-parameter 
models integrating social factors into link prediction. 

Models with pure network-structure prediction assumptions derive from the vast research on global network structures. Studies on network properties 
confirm that many real-world networks display small-world properties in which high node clustering is combined with short average internode 
distances~\cite{ws98,newman2004cna}. Networks have also been found to behave according to a power-law scale-free phenomenon where 
a relatively small number of nodes have numerous connections~\cite{price1965,barbasi02,akke:2012}. Additionally, 
networks have properties of clustering hierarchies~\cite{barbasi02}, and tendencies of transitivity or ``triangles 
of interaction'' describing the manner in which ties between node $A$ and $B$, and between node $B$ and $C$ facilitate 
a likely tie between $A$ and $C$.

From this knowledge on network properties a second generation of studies emerged addressing how a social network can be modeled using properties intrinsic to the network. 
These global network-structure link-prediction models provide insight into not yet identified or observed linkages \cite{krebs2002}, as well as to infer not directly 
observed likely links \cite{goldbergroth2003,popescul2003,taskarwong2003}. Within these studies two approaches are taken to predict links: 
(1) computing node-level measures from greater network structures and, 
(2) meta-level analyses. 
In this study we consider only node-level measures (which are comparable to the gap we aim to fill in this research), while still maintaining the 
network structure. 

Several approaches for predicting social network linkages have been proposed, for a complete list see 
\cite{dln}. Despite the extensive research of different measures used to model the network dynamics, all of these models suffer from low fitness, 
with random link prediction performing just as well as Katz's model of path collection- predicting links by the sum of collected path lengths per individual \cite{Katz94}. This has led informaticians to explore the 
effects of additional parameters in understanding network dynamics. 
A second model type works to address the effect(s) of social parameters on the dynamics of social networks. The justification for these models 
arose from research on social networks which proved that social selection plays a key role in relation formation~\cite{granov73,krack92,enne:baum:1994}. 
Models of this type allow us to question how a social network can be modeled using both network and social properties of nodes. These models also 
infer edges through evolving graph models but consider state spaces with both network and social parameters. Two model types are commonly used to 
investigate the inference of these dual parameters: stochastic actor models (SIENA)~\cite{snijders:2010} 
and exponential random graph models (ERGM)~\cite{pstar}.   

The key distinction in these models, from the network-only models, is the combination of link prediction based on both local effects, as well as 
on ``social circuits'' that capture the influence of more distant ties on behavior \cite{robins2011}. This leads to an exponential growth of the state space due to the consideration of more parameters, requiring extensive computing power in prediction. Given the computational complexity of calculating this for every 
node these models are not easy to develop in a way that convergence emerges in large networks \cite{robins2011}.  Consequently, these classes 
often limit the size of networks through a theoretical boundary of inferring statistics for a bounded network. This reduces the burden of having 
to perform computations on potentially very large graphs, but also effectively limits application to small networks ($\sim$ 1000 nodes).

In summary, these two model classes provide a testing ground to explore dynamics, but are both not without limitations. Both network and social 
parameters have scalability problems. As we discuss next, in order to empirically explore the effect of both network and social parameters on 
large social network dynamics a scalable solution is required.

\section{Modeling Framework}
\label{sec:model}
We propose a \emph{mean-field approach} for studying social networks; (equally behaving) individual nodes are grouped according to their 
\emph{states}. This approach is used for an optimized analysis of large-scale systems, allowing for a prediction of the average behavior of 
the system. The mean-field theory has been applied previously, e.g., to large-scale gossip systems in~\cite{per10,BEEFH10,rena-thesis}. 
Concisely, the state of the system is represented by a distribution, or a vector of fractions of nodes $\delta_s(t)$ in each state $s$ 
at time unit $t$. The evolution of the stochastic system is governed by a so-called master equation of the form: 
{
\begin{align}\label{eq:mastereq}
\delta (t+1) = M_{\delta(t)} \cdot \delta (t)
\end{align}}%
$M_{\delta(t)}$ is the matrix, each entry of which is a transition probability from a state $s$ at time $t$ to state $s'$ at time $t+1$.
Thus, we are effectively reducing the global state space, thereby increasing the computational efficiency of the model, and in turn, allowing us to consider more parameters as well as more nodes.

Moreover, we use the notion of \emph{classes}, introduced in~\cite{BEEFH10}, to distinguish between equally behaving 
nodes affiliated to different categories. To this end, the mean-field model predicts average behavior 
of sets of nodes of each class given a number of social and network parameters. 
We highlight the modelling steps:

\paragraph{Forming a model}
In order to model the network, first we need to define the system in the form of its parameters. This will form a state of the system. Given the type of network under study, the effects of system parameters are considered using either manual classification or statistical classification (e.g., \cite{Bis95}) to identify the set of significant parameters to form states and classes. For example, some parameter $u$ can be a theoretically informed organizational constraint (e.g. an organization, a background, etc).

\paragraph{Applying abstraction refinement}
The theory underlying the mean-field model requires also the population of each state to be large enough to be approximated by the law of large numbers. The size of the population in a sampled data set may force one to consider further abstraction for the ranges of the parameters, thereby reducing the size of the system state space. For instance, if chosen parameters for the system are the number of papers per author $p \in \mathbb{N}$ and the number of an author's coauthors $c \in \mathbb{N}$, the number of possible states of the system will simply be a product $\mathbb{N} \times \mathbb{N}$.
Some parameters can be restricted in their value ranges without loss of the accuracy of the model itself.

\paragraph{Computing the model input}
To execute the model, input data is needed on the initial state of the system, as well as on distributions for networking behavior, which will 
be used for the matrix $M_{\delta(t)}$. The input distributions for the mean-field model include three categories: (1) communication, (2) idle, and  
(3) collision. Communication describes the interaction between nodes, and idle is a state of no interaction. Collision is the disappearance or 
decay of an interaction. The distributions of interaction (links, from a graph-theoretical perspective) are estimated for each class, 
which determines the nonuniform behavior by different classes for the model. We compute these distributions statistically from 
the sampled data set. 

\paragraph{Estimation of distributions}
The aforementioned transition probability distributions are determined using a discrete-time model to identify
the optimal time slicing for the studied data set. Such a time slice corresponds to one time unit in the model. 
The distribution for probability of transition from one class to another one is also used in the master equation 
\eqref{eq:mastereq} (for a more detailed equation, cf.~{\cite[Fig.~10]{BEEFH10}}. 
The method used for estimation of the probability distributions is a Hidden Markov Model (HMM)~\cite{hmm}. 

\paragraph{Applying automated mean-field framework}
Armed with the knowledge regarding states, classes and transition rates, obtained from the previous steps, we apply an automated mean-field framework to infer average behavior of the system. 
We repeat the earlier steps until all parameters are included for a time period covered by the data set. 
We use the resulting mean-field model to make average link predictions on the system given the parameters under consideration. The model provides a number of advantages over models discussed in Section~\ref{sec:models}, such as greater flexibility in modeling behavior of nodes through a number of processes. The use of HMMs provides an additional round of probability in node interactions, to compensate for the aggregation. Moreover, such a model allows us to consider both social parameters as well as network structures. Unlike simulation or deployed models, the model is flexible given a theoretical knowledge of the interactions under study.  In analyzing the system under question we set the formal specifications which provide detailed processes of specification.

\paragraph{Considerations for extensions of social networks} 
The challenge in applying the mean-field model to social networks is to derive accurate predictions of the local 
behavior of the nodes within defined classes. Particularly, for social networks, model abstractions need to be 
done using a data-aware approach. A data-aware approach implies that both classes and parameters are informed 
through an intense, robust knowledge of the system under study, as well as the content of edges in the network 
data. It is a requirement that this is approachable through a theoretically or empirically grounded conceptual scheme on both the system under study and the mechanisms that inform the parameters considered in simulation models. Consequently, not all social 
networks and or systems can be analyzed using such an approach.  

Additionally, we argue for an interdisciplinary approach in development of the model as data needs to be 
intensely explored to inform parameters by both a data engineer and validated by social scientists or 
informed experts of the system under study. This implies, unlike other models, that the data-aware approach 
is essential to determining accurate results, which can be compared in model-fit tests. This results in a model 
that specifically fits the needs of the system under study, and which can be adapted per population given 
the basic set of rules for abstraction we describe.
In the next section we lay out the general steps for the application of a mean-field model. 

\section{Application}
\label{sec:Application}
As discussed in the previous section a set of requirements are necessary for 
implementing a mean-field model to investigate the effect of social and network factors on network 
dynamics: network data, parameter data, and knowledge from empirical studies of the system under study. We 
explain the case studied here and detail the abstraction steps undertaken to model the effect of network and social parameters 
on network dynamics.

\subsection{Network data} 
A majority of computational analyses of large social networks implement coauthor or similar co-occurrence 
networks to examine network dynamics \cite{barbasi02}. Coauthorship networks, via publication data, provide a representation of a specific social interaction- successful collaboration, in producing an output- dissemination of knowledge through publication. Moreover, 
publication data is readily accessible on the Web providing large, reliable, and scalable data sets to model 
network dynamics. 

In addition to the use of coauthorship data to study network dynamics, empirical studies on coauthorship 
provide a framework to develop measures to consider in the model testing. In science studies, coauthorship 
is a standard measure for collaboration in science. Collaboration is increasingly common in science; from 
the near disappearance of single-authored papers to the growth in prevalence of an increasing numbers of 
coauthors on academic publications \cite{grenee2007}. 
A decade of studies on collaboration in 
science have proved the effect of different social variables on collaborative behavior of scientists 
\cite{bozemancorely2004,stokols2008}. Recent studies have found that task types and a number of external 
factors influence collaborative behavior of scientific processes \cite{borner10}. Both institutional and 
short geographical distances play a key role in the collaborative behavior of scientists \cite{rodpepe09,uzzi08}. 
Given these studies we have a basis at which to both test informed parameters and link findings to 
knowledge on collaborative tendencies of scientists. 

In this paper we explore a system of collaborative behavior of scientists in testing the mean-field model 
for large social networks. We select one nation and discipline -- Dutch computer scientists, to investigate 
dynamics as to limit known exogenous effects of different knowledge production practices between disciplines 
and nations. Effectively, we comment only on the average behavior of the system of Dutch CS.
The field of CS was chosen for three reasons: the traditions of the field with a diversity of 
subfields within the discipline; the known tendency for collaboration through coauthorship; the validity 
and reliability of online sources documenting publications. The Dutch context provides a diversity of cases at which to examine different institutional processes.  

A source list of 434 tenured Dutch computer scientists in 2010 was acquired from the Nederlands Onderzoekdatabank, 
an official body that keeps records on research in the Netherlands. To identify a valid and reliable set of 
coauthorship data for the Dutch computer scientists a snapshot of DBLP DataBase 
was queried. (DBLP is one of the most comprehensive bibliographic indices for the field of CS.)
%
Within this set the list of Dutch computer scientists was queried for all publications of scientists from 2006 - 2010 
(the year of our list of tenured scientists). This list was manually cleaned to disambiguate names. From this list the name of the 
publication was queried to identify the unique author IDs of each author per publication. These unique author IDs were queried 
to pull full publication lists of each author (Dutch scientists and their coauthors). 

Conference proceedings were selected for the case study as conferences in CS require at least 
one author to physically present work at a conference to be published. Conferences provide a good fit for the assumption of interaction 
in previous computer models as a potential meeting points for coauthors. Additionally, it provides a number of 
clear timestamps discerning possible transition periods, with most conferences occuring annualy, with regular cycles. Conference 
proceedings are denoted in this data set by the BibTeX entry \texttt{@inproceedings}, allowing us to further query for proceedings-only 
publications. This resulted in 3639 scientists, and 2757 conference-proceeding publications. Nodes represent individual scientists and 
links represent shared coauthorship of proceedings. From this data set of individual authors we also collect data on the social parameters. 

\subsection{Parameters}

In this study we aim to include parameters that are informed from previous empirical studies in the field of science studies. Four parameters 
are considered in the model: scientific age, cosmopolitanism of knowledge production, visibility, and institutional affiliation. For the collection 
of social parameter data in this study the Web is used, providing a reliable method for collecting meta-data on scientists within publication 
records~\cite{mika06}. The use of Web data as the source of meta data is integral in this first model development as it reduces the burden of data 
collection of social variables (compared to traditional social science data of surveys or interviews). This allows us to quickly test the effect 
of social parameters on behavior with a considerable amount of reliability from merging meta-data from additional online databases.

The parameters -- scientific age, cosmopolitanism of knowledge production, and visibility are calculated from within the DBLP data set. Scientific 
age was selected because tenure and rank are both said to play a role in collaborative behavior of scientists, with scientists of a higher tenure 
more likely to collaborate than mid-range, tenure-seeking colleagues \cite{beaverrosen1979}. We first noted publication per author in the DBLP 
data set for which we compute per year per author as his or her scientific age. A second parameter, cosmopolitanism, relates to the socio-technical 
acquired capabilities of scientists suggesting that access to potential coauthors in a field plays a key role in collaboration~\cite{bozemancorely2004}. 
This parameter was measured through previous coauthorship experience. The number of coauthors per year per author is computed from the DBLP. The 
third parameter aims to comment on the visibility of the scientist. The visibility of 
the scientist is the likely popularity through publication magnitude. These three parameters allow us to consider a number of possible social 
factors that are not network effects but rather social attributes on the scientists' networking behavior.

One additional parameter was collected for consideration in the model -- the institution.  Previous studies proved that 
the institution is statistically significant with respect to how scientists collaborate \cite{rodpepe09,uzzi08,borner10}. The institution is identified through a query of two databases. These data are considered static in this 
model, unlike the previously mentioned data, as we assume minimal change of institution in the five-year period 
under study. The automatic collection of historical data on institutional affiliation is not currently stored in one 
database, to our knowledge, thus we assume a five-year period as a valid period of time to accurately measure 
inference. A query using Microsoft Academic Search -- a database which includes the DBLP data set is used to 
identify institutions. To locate additional missing data another database, ArnetMiner.org was used.
The remaining unidentified institutions were queried manually giving us a total of 1358 identified institutions. In order to disambiguate 
institutional names, to have a reliable and valid set of data, this list was queried in geocoding Web service Yahoo! 
PlaceFinder~\cite{yahoo}. 
This query provides a proximity measure for each institution and a uniform institutional affiliation based on 
common GPS coordinates. 

These four parameters provide a setting to explore the application of the mean-field model in large social networks. The occupancy measure at time 
$\delta(t)$ in our model is the fraction of people in state $(p, c, h, u)$, where $p$ is a number of publications, $c$ is a number of coauthors, 
$h$ is scientific age, and $u$ is affiliation. We test the following social science hypothesis: institutions effect the patterns of collaborative behavior 
(by behavior we mean average number of coauthors, and average number of papers).
In addition to these social parameters we also include the network parameter of transitivity. As discussed in section 1, social networks have tendencies of transitivity \cite{barbasi02,newman2004cna}. We consider the social parameters in predicting the triadic interactions between nodes. 

\subsection{Classes abstraction}
  
In principle, any of our parameters could be considered a class. When studying a social system, however, we need to consider known social 
and organizational constraints. In order to define a class we investigate the four possible parameters under consideration in this model. 
We first consider known effects. 

Our system is already bounded by the selection of one national science structure and one scientific discipline. The effect of the 
institution provides a valid and logical boundary at which to explore aggregation. Additionally, we know that geographical location also plays 
a key role in collaboration, which we aim to consider in the abstraction. Consequently, we employ institutions as classes in our mean-field 
model, and as one of the parameters $u$ contributing to a state $(p,c,h,u)$ of a collaboration network. Due to limitation of the data-mining 
techniques to automatically extract full history of  scientific employment, we assume that a scientist has one affiliation during the four year period. 

The data set for our model consist of 3639 Dutch authors with 749 different institutions. However, the theory underlying our mean-field model 
requires that the population of each class should be large enough to be approximated by the law of large numbers. To this end, we applied an 
abstraction on classes (institutions) based on statistical metrics for the given distribution $D$ of computer scientists among institutions. 
  
Since both our data set and results are focused on the system of Dutch computer scientists, we distinguish (1) institutions in the Netherlands, 
and (2) institutions in other countries. For each of these categories we estimate a statistical threshold of the significance of the institution.  
This threshold depends on the dispersion of the distribution $D'$ of scientists sampled for each of the categories of institutions. If values 
are highly dispersed, then we set the threshold to be the average number of affiliated scientists.
   
To measure the statistical dispersion for the scientists' distribution $S$, we compute \emph{a sample covariance}, which is the average distance 
to the mean value between any two values in the distribution $S$. To allow for some dispersion, we compare the arithmetic mean for $S$ and its 
sample covariance: if the sample covariance for a subset $S \in D$ is higher than the mean, then the values of the sampled $D'$ are highly 
dispersed. 
   
   \begin{figure*}[b!]
    \vspace*{-0.2cm}
  \begin{minipage}[c]{0.51\linewidth}
  \includegraphics[scale=0.35]{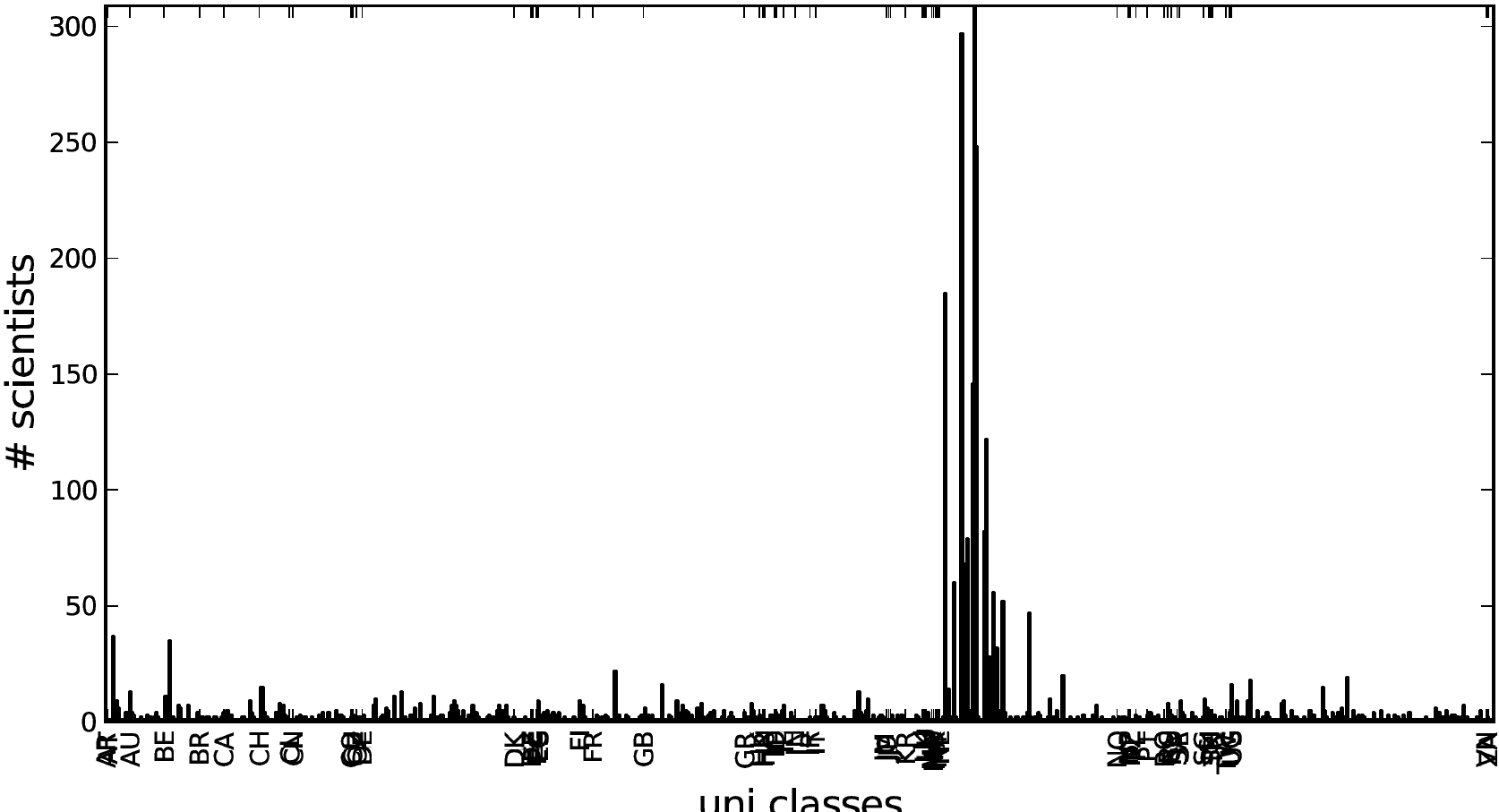}
  \end{minipage}
  \begin{minipage}[c]{0.4\linewidth}
   \includegraphics[scale=0.35]{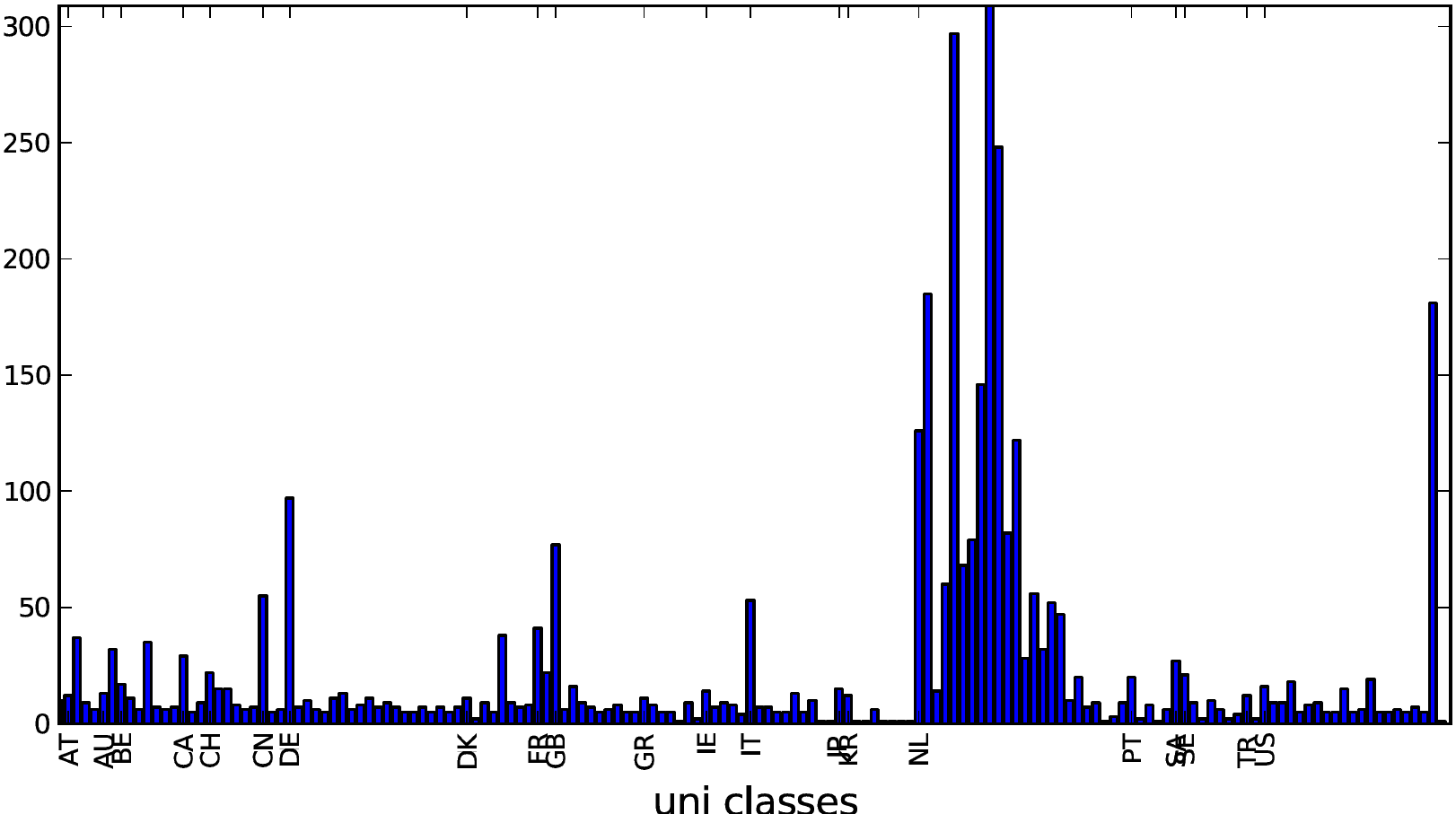}
  \end{minipage}
  \caption{The distribution of scientists among institutions before (left) and after (right) the abstaction}
  \label{fig:histo}
  \end{figure*} 

In addition to estimation of the significance threshold, this simple test is applied in two steps: (1) for 
the continental abstraction, and (2) the country-wide abstraction. In case 1, we sample data for all 
universities per continent (using the UN list of countries per continent and GPS coordinates). In the case of 
high dispersion in the number of scientists in institutions in one continent, we proceed to test the dispersion 
of the number of scientists affiliated with institutions in one country. We merge only those institutions that 
have a number of scientists below the mean of the entire distribution $D$. The histogram in Fig.~\ref{fig:histo} 
shows the number of scientists in each class, before and after the classes abstraction. The number of classes has 
been reduced from an initial 749 to 157, effectively reducing also the state-space size.

 \subsection{Other parameters abstraction}  
   \paragraph*{Scientific age} 
   The scientific age $h$ is based on the first publication 
   date of an author according to DBLP. The earliest possible publications in DBLP date back to 1971, which inevitably leads to 
   an increase by a factor 40 of the state-space size of our model. Considering our sampled data set with only 3639 scientists, 
   the distribution of the population in such a state space is very sparse. Thus, we identify five main groups of scientific age, 
   categorizing age into ten-year periods as to generalize about generations of scientists: 
   $70$, $80$, $90$, $2000$, $2010$. In general, scientific careers require substantial 
   investments to establish tenure. These positional differences, whether it 
   being established tenure, or a starting PhD, all influence the manner in which scientists undertake collaboration ~\cite{beaverrosen1979,bozemancorely2004}. 
   Our abstraction granularity is fine enough to strongly indicate the scientific position of researchers, e.g., senior staff, 
   junior staff.
   \paragraph*{Visibility}
   The visibility of the scientists is measured by the annual number of conference publications. We choose only conference 
   publications, as a potential interaction point, assuming that scientists encounter future collaborators during conferences. 
   Without loss of generality, we limit the highest number of conference publications per year to 12 assuming it takes on average one month of preparation per publication. Those scientists 
   that publish 12 and more papers per year we distinguish as fast publishers with a parameter value of $12$.
   \paragraph*{Cosmopolitanism}
   The cosmopolitanism of the science is measured by number of coauthors, indicating how well connected a scientist is. We studied the distribution of the number of coauthors on our sampled data set. We observed that there are few publications with a large (more than 12) number of coauthors on a single 
   paper. A high number of coauthors on a paper generally indicates a participation in a large research project. This results in an unnecessary large 
   state-space size of the model, given the sampled authors in this sample. To tackle this, we distinguish five categories of  
   coauthor count per paper: ``non cooperative'' ($0$) for the papers with one author, ``regular'' ($1$) for the papers 
   with up to $3$ coauthors, ``high'' ($2$) with up to $6$ coauthors on the paper, ``team'' ($3$) with up to $10$ coauthors, 
   and a ``large project'' ($4$) for papers with more than $10$ coauthors. Since we consider the unique 
   coauthors of a scientist as possible network contacts within one year, we take the annual number of coauthors 
   relative to the number of the publications per year per person.

   \subsection{Transitions and Distributions}
   
   There are three categories of distributions needed to derive from our data set for our mean-field model: 
   (1) communication $\kappa$, (2) idle $\eta$, and (3) collision $\phi$. \emph{Communication} is defined as 
   collaboration via shared coauthorship between two scientists resulting in a conference paper. 
   Both \emph{idle} and \emph{collision} states signify the decay of communication; in fact, 
   for our application, these probability distributions are both an identity function. Moreover, in terms of the model, selection of the collaboration partner is governed by the distribution 
   function $\textit{contact}$, which specifies the collaboration network topology. 

   \paragraph{Computing transition probabilities} We first measure from the collected data the evolution of collaboration 
   between scientists (nodes) for each year 2006--2010. That is, we compute the state vector $\delta(t)$, entries of 
   which are the fractions of nodes in every possible state of the system at time $t$. This state vector $\delta(t)$ is 
   used in the initial configuration for the model: 
   we sum up all fraction of nodes with scientific age $h$ from class $u$, $\delta_{(p,c,h,u)}(t)$ for all possible $p$ and $c$ 
   and set the result as $\delta_{(0,0,h,u)}(0)$ at the beginning of each year $t$. In the model, we split the time frame onto 
   a week $\tau$, for finer granularity, with 52 weeks in each year.   

   Consider states $A = (p_a,c_a,h_a,u_a)$ and $B = (p_b,c_b,h_b,u_b)$. For each pair of classes $u_a$ and $u_b$, we compute the probability 
   $\textit{contact}(u_a,u_b)$ that a node from $u_a$ contacts any node in $u_b$ in year $t$ as follows. Each paper $i$ with $c_i$-authors 
   by a node from $u_a$ and a node from $u_b$ gives the probability $P_i(c_i,u_a,u_b) = \frac{1}{{m}(u_a) \cdot c_i}$ that the node from 
   class $u_a$ contacts a node  
   from $u_b$. Here, ${m}(u_a)$ is the number of nodes in class $u_a$. Since we have to take into account that papers jointly written 
   by nodes from $u_a$ and $u_b$ may have other coauthors, divisor $c$ distributes the share of contribution to each coauthor. 
   Then, $\textit{contact}(u_a,u_b)(t)$ is obtained as follows: $\textit{contact}(u_a,u_b)(t) = \sum_{i(u_a)} \sum_{i(u_b)} P_i(c_i,u_a,u_b)$, 
where $i(u_a)$ and $i(u_b)$ means ``for each author of paper $i$ from class $u_a$'' ($u_b$, respectively).

   The computation of the collaboration distribution $\kappa_{(A,B)}(t)$ is as follows. For each paper penned by authors in states $A$ 
   and $B$ (within a one-year time frame), we observe all possible state transitions (i.e. before and after collaboration). The result is  
   an expression of the form:
   \begin{align*}\label{eq:kappa}
     \hspace*{-0.3cm}
  \kappa_{(A,B)}(t) =  \{(p_1,(A,B),(A_1,B_1)), \ldots (p_n,(A,B),(A_n,B_n))\}
       \hspace*{-0.3cm}
   \end{align*}
where $p_{i}$ is the probability that the nodes in state $A$ at time $t$ make a transition to state $A_i$ at time $t+1$ 
(and, those in state $B$ move to state $B_i$, respectively).
All these distributions are normalized to a weekly timescale. 
  
 \paragraph{Estimating distributions}

   These rates may vary from year to year thereby requiring an average to be determined for every of these distributions to ensure accuracy in 
   the model. To that end, we obtained probabilities, as described earlier, for the years 2006--2008, and use an HMM approach to sample 
   the underlying distribution.\comment{ We can view ``training-period'' collaborations as samples drawn from a probability distribution on pairs of nodes.}
   Our goal is to approximate the set of pairs that have positive probability of collaborating. Our mean-field model takes these sampled distributions as its input.

\section{Results}

\begin{figure*}[b!]
  \centering
  \subfloat[Comparison with mean-field.]{\label{fig:mfvs}\includegraphics[width=0.48\textwidth]{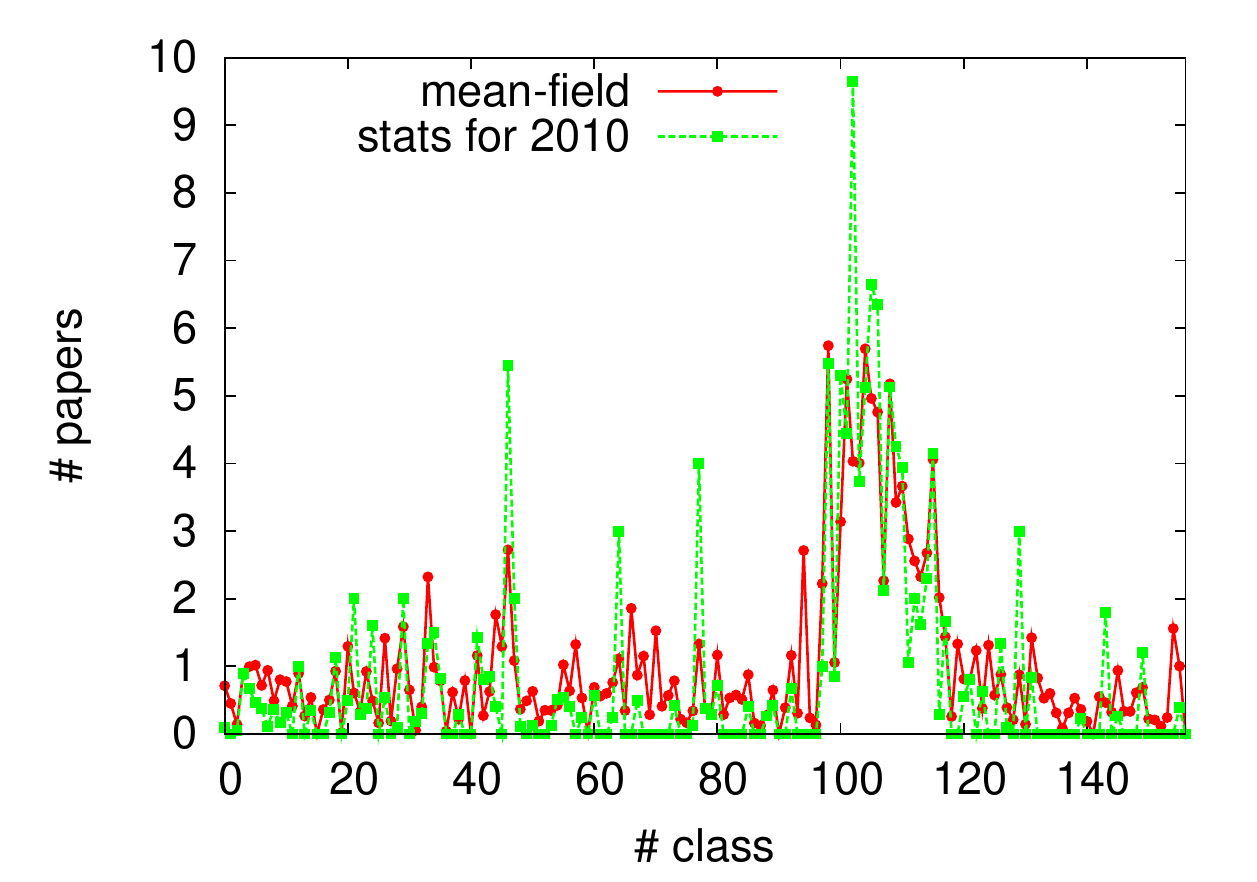}}
  \subfloat[Comparison of different $\textit{contact}$.]{\label{fig:uniform}\includegraphics[width=0.48\textwidth]{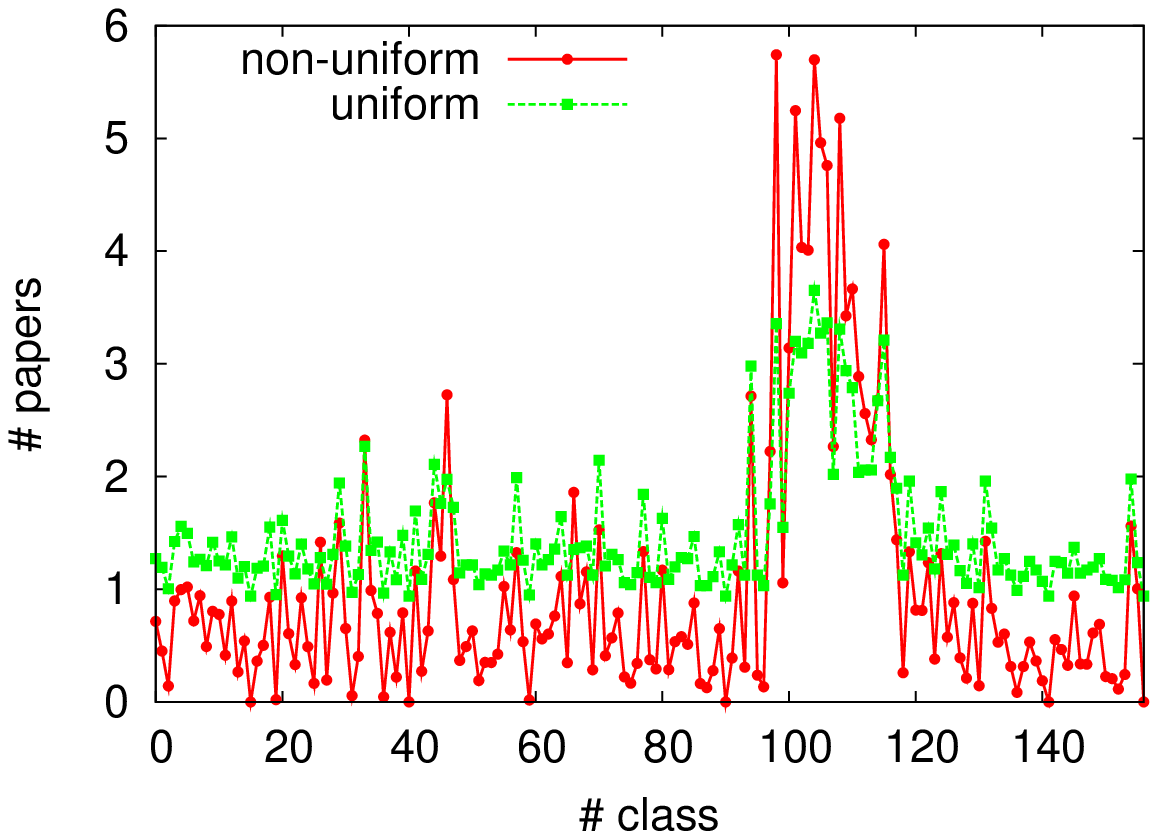}}
  \caption{Average output for different classes.}
\end{figure*}

The mean-field model allows us to predict average behavior. The analytical results to the statistical results for the years 2009 and 2010 are compared to the ones produced by the mean-field model.
Institutions are labeled and sorted in lexicographical order; this list is enumerated and corresponds to the number on the $x$-axis 
(similar to Fig.~\ref{fig:histo}). Classes 98-116 correspond to Dutch institutions. 
As we can see from Fig.~\ref{fig:mfvs} the mean-field results for the larger institutions corresponds with the statistics from the data set 
for 2010. Our data set does not list all papers of the coauthors of coauthors, but we divide by all people in the class;
so statistics produced are lower than actual.

\paragraph{Institutional factor}

The results produced by the alternative mean-field model with uniform distribution $\textit{contact}$ for collaborations between 
different institutions show that the sample distribution is non uniform. This $\textit{contact}$ distribution produces the equal 
probability of collaboration between any two scientists in the whole network, irrespective their affiliations, and thus forms a baseline for comparison to see whether affiliations are statistically significant. The comparison in shown in Fig.~\ref{fig:uniform}.
As we can see, the uniform $\textit{contact}$ 
distribution predicts higher output for foreign institutions but lower for Dutch institutions, since the output is then uniformly ``redistributed''.



\paragraph{Impact of scientific age}

\begin{figure*}[!th]
  \centering
  \subfloat[Average output for different scientific age.]{\label{tab:age}
    \setlength{\arrayrulewidth}{0.85pt}\renewcommand{\arraystretch}{1.2}
  \begin{tabular}{c|c}
    \centering
    Sci. age  & avg \# pubs. \\
    \hline
    2010s &  1.8 \\
    2000s &  1.61 \\
    1990s &  1.76 \\
    1980s &  1.95 \\
    1970s &  2.3 
  \end{tabular}    
    }{\hspace{0.2cm}}
  \subfloat[Triad relations.]{\label{tab:triad}
\begin{tabular}{c|c}
    $u_a \leftrightarrow u_b$, $u_b \leftrightarrow u_c$ & avg. $u_a \leftrightarrow u_c$ \\
    \hline
     $>=$ 0.0 & 1.0\\
     $>=$ 0.2 & 1.13\\
     $>=$ 0.4 & 1.15\\
     $>=$ 0.6 & 1.20\\
     $>=$ 0.8 & 1.27\\
     $>=$ 1.0 & 1.32
    \end{tabular}    
    }
\caption{Results for the age impact and triad relations for Dutch institutions.}
\end{figure*}

Fig.~\ref{tab:age} shows the average number of papers for different scientific age. The results from only Dutch institutions were averaged.
The mean-model model shows that a principle of preferential attachment \cite{barbasi02} is occuring in the network based on age, with higher 
tenured scientists acquiring more collaborators and papers. The average output per scientific age per institution, was also computed; see 
results\comment{ in Table~\ref{tab:age-uni}} in~\trref, which displayed differing tendencies in collaboration patterns.

\paragraph{Link prediction}
 
In accessing the manner in which links are made through transitivity: if class $A$ has a paper in common with $B$, and class $B$ with $C$, 
then $A$ has stronger connectivity with $C$. Within this system we consider the institution parameter, allowing us to reflect on the initial hypothesis -- an institution plays a role in the collaborative patterns of scientists. The connectivity factor based on the distribution $\textit{contact}$,
which in turn, depends on the probability $P_i(c_i, u_a, u_b)$, 
the number of coauthors from a certain institution implicitly contributes to strength of the connectivity between institutions. 
Fig.~\ref{tab:triad} shows the generalized triad relations of Dutch institutions; considering a scientific age in $\textit{contact}$ produces 
results\comment{ in Table~\ref{sec:relations}} in~\trref.

\section{Discussion and Conclusion}

In investigating the system of Dutch computer scientists' collaborative behavior through the mean-field model we observed 
systematic networking behavior associated with a number of social parameters, which aid in describing the networking dynamics 
of scientists. The past collaborative partners of one's institution plays a key role in how future collaborations unfold. 
With every conference proceeding with another institution the chance of collaborating with the institution increases.  
Age also matters; the age of the scientists plays a role in the visibility of a scientist (number of publications) within the system. 
The cosmopolitanism of the scientists (number of co-authors) also contributes to the likelihood of future interaction. Consequently 
the mean-field model allows us to describe the Dutch CS system of conference paper collaboration to be governed by a 
number of social variables, where ties can be predicted given previous relationships among common institutions, 
reinforcing clustering tendencies in these networks.

In this first application of the mean-field model in predicting both social and network parameters for large social 
networks, we also recognize a number of shortcomings. The first is the sensitivity of the data-aware approach and thus 
the empirically informed aggregations of nodes into clusters from such an approach. Future work should aim to consider 
additional social parameters, such as performance, gender, discipline, length of time known in understanding the 
system. To improve the precise description of states the notion of idle and collisions in the model should be improved 
for social networks. Additionally, we acknowledge that this explorative study of the mean-field model did not address 
both the potential for shift classes reflecting the fluidity of actual organization constraints in social life, as well 
as model checking. These limitations are related to the current state of computing techniques, in first data-mining 
techniques which does not currently allow us to collect such refined information on social beings, and secondly the 
lack of methods to appropriate accurate model checking.

The incorporation of the modeling knowledge with population specific dynamics we are able to identify the conditions under which links emerge 
given a set of both network and social parameters through the mean-field model. This allows us to provide  informed predictions to comment on 
the mechanism(s) under which specific patterns of behavior emerge in large social networks. Mean-field models provide a meta-scopic method, 
which overcomes limitations of the network only and social parameter models. Meta-scopic models of this sort allow us to incorporate both the micro (considered in evolving 
graph models) and the mega networking processes to infer links through a data-aware approach. Additionally, it provides an empirical terrain 
at which to explore the effects of both network and social parameters on large social networks. 

\subsection*{Acknowledgements}
We thank Paul T. Groth for the initial DBLP data set, and J\"{o}rg and Stefan Endrullis for their support in the "refitting" of the automated mean-field framework for the social domain.
\bibliographystyle{splncs}

\begin{thebibliography}{10}

\bibitem{Ahuja2012}
Ahuja, G., Soda, G., Zaheer, A.:
\newblock The genesis and dynamics of organizational networks.
\newblock Organization Science \textbf{23} (2012)  434--448

\bibitem{snijders:2010}
Snijders, T., van~de Bunt, G., Steglich, C.:
\newblock Introduction to actor-based models for network dynamics.
\newblock Soc. Networks \textbf{32} (2010)  44--60

\bibitem{barbasi02}
Albert, R., Barab\'{a}si, A.L.:
\newblock Statistical mechanics of complex networks.
\newblock Reviews of Modern Physics (74) (2002)  47--97

\bibitem{barabasialbert1999}
Barab\'{a}si, A.L., Albert, R.:
\newblock Emergence of scaling in random networks.
\newblock Science \textbf{286} (1999)  509--512

\bibitem{dln}
Liben-Nowell, D., Kleinberg, J.:
\newblock The link-prediction problem for social networks.
\newblock J. ASIST \textbf{58}(7) (2007)  1019--1031

\bibitem{newman06}
Moore, C., Ghoshal, G., Newman, M.E.J.:
\newblock Exact solutions for models of evolving networks with addition and
  deletion of nodes.
\newblock Phys. Rev. E \textbf{74} (2006)  036121

\bibitem{newman2004cna}
Newman, M.:
\newblock Coauthorship networks and patterns of scientific collaboration.
\newblock Proc. Natl. Acad. Sci \textbf{101} (2004)  5200--5205

\bibitem{Barabasi2002590}
Barab\'{a}si, A., Jeong, H., N\'{e}da, Z., Ravasz, E., Schubert, A., Vicsek,
  T.:
\newblock Evolution of the social network of scientific collaborations.
\newblock Physica A: Statistical Mechanics and its Applications
  \textbf{311}(3--4) (2002)  590 -- 614

\bibitem{grossman:03}
Grossman, J.W.:
\newblock Patterns of collaboration in mathematical research.
\newblock Notices of the {AMS} \textbf{52}(1) (2005)  35--41

\bibitem{ws98}
Watts, D.J., Strogatz, S.H.:
\newblock Collective dynamics of `small-world' networks.
\newblock Nature \textbf{393}(49) (1998)  440--442

\bibitem{price1965}
de~Solla~Price, D.:
\newblock Introduction to the special issue on network dynamics.
\newblock Science \textbf{149}(3683) (1965)  510--515

\bibitem{akke:2012}
Akkermans, H.:
\newblock Web dynamics as a random walk: How and why power laws occur.
\newblock In: Proc. Conf. of Web Science (WebSci), ACM (2012) To appear.

\bibitem{krebs2002}
Krebs, V.:
\newblock Mapping networks of terrorist cells.
\newblock Connections \textbf{24}(2) (2002)  43--52

\bibitem{goldbergroth2003}
Goldberg, D., Roth, F.:
\newblock Assessing experimentally derived interactions in a small world.
\newblock In: Proc. Natl. Acad. Sci. (2003)  4372--4376

\bibitem{popescul2003}
Popescul, A., Ungar, L.:
\newblock Statistical relational learning for link prediction.
\newblock In: Proc. Conf. on Artificial Intelligence, ACM (2003)  81--90

\bibitem{taskarwong2003}
Taskar, B., Wong, M.F., Abbeel, P., Koller, D.:
\newblock Link prediction in relational data.
\newblock In: Proc. of Neural Information Processing Systems, MIT Press (2003)
  659--666

\bibitem{Katz94}
Katz, L.:
\newblock A new status index derived from sociometric analysis.
\newblock Psychometrika \textbf{18}(1) (1953)  39--43

\bibitem{granov73}
Granovetter, M.:
\newblock The strength of weak ties.
\newblock American Sociological Review \textbf{78} (1973)  1360--1380

\bibitem{krack92}
Krackhardt, D.:
\newblock The strength of strong ties: the importance of philos in
  organizations.
\newblock Netw. and Organiz.: Structure, Form, and Action (1992)  216--239

\bibitem{enne:baum:1994}
Ennett, S., Bauman, K.:
\newblock The contribution of influence and selection to adolescent peer group
  homogeneity, the case of adolescent cigarette smoking.
\newblock J. of Personality and Social Psychology \textbf{67} (1994)  653 --
  663

\bibitem{pstar}
Robins, G., Pattison, P., Kalish, Y., Lusher, D.:
\newblock An introduction to exponential random graph (p*) models for social
  networks.
\newblock Soc. Networks \textbf{29}(2) (2007)  173--191

\bibitem{robins2011}
Robins, G.:
\newblock Exponential random graph models for social networksl.
\newblock In: Handbook of Social Network Analysis.
\newblock Sage (2011)

\bibitem{per10}
Bakhshi, R., Cloth, L., Fokkink, W., Haverkort, B.R.:
\newblock Mean-field framework for performance evaluation of push-pull gossip
  protocols.
\newblock Performance Evaluation \textbf{68}(2) (2011)  157 -- 179

\bibitem{BEEFH10}
Bakhshi, R., Endrullis, J., Endrullis, S., Fokkink, W., Haverkort, B.:
\newblock Automating the mean-field method for large dynamic gossip networks.
\newblock In: Proc. of QEST, IEEE Computer Society (2010)  241--250

\bibitem{rena-thesis}
Bakhshi, R.:
\newblock {Gossiping models: Formal Analysis of Epidemic Protocols}.
\newblock PhD thesis, Vrije Universiteit Amsterdam (2011)

\bibitem{Bis95}
Bishop, C.M.:
\newblock Neural Networks for Pattern Recognition.
\newblock OUP (1995)

\bibitem{hmm}
Stratonovich, R.:
\newblock Conditional markov processes.
\newblock Theory of Probability and its Applications \textbf{5} (1960)
  156--178

\bibitem{grenee2007}
Grenne, M.:
\newblock The demise of the lone author.
\newblock Nature \textbf{450}(1165) (2007)

\bibitem{bozemancorely2004}
Bozeman, B., Crley, E.:
\newblock Scientists' collaboration strategies: Implications for scientific and
  technical human capital.
\newblock Research Policy \textbf{33}(4) (2004)  599--616

\bibitem{stokols2008}
Stokols, D., Misra, S., Moser, R., Hall, K., Taylor, B.:
\newblock The ecology of team science: understanding contextual influences on
  transdisciplinary collaboration.
\newblock American Journal Preventive Med \textbf{35}(2S) (2008)  S96--S115

\bibitem{borner10}
B\"{o}rner, K., Contractor, N., Falk-Krzesinski, H.J., Fiore, S.M., Hall, K.L.,
  Keyton, J., Spring, B., Stokols, D., Trochim, W., Uzzi, B.:
\newblock {Team Assembly Mechanisms Determine Collaboration Network Structure
  and Team Performance}.
\newblock Sci. Transl. Med. \textbf{2}(49) (2010)  49cm24

\bibitem{rodpepe09}
Rodriguez, M., Pepe, A.:
\newblock On the relationship between the structural and socioacademic
  communities of a coauthorship network.
\newblock J. Informetrics \textbf{2}(3) (2009)  195--201

\bibitem{uzzi08}
Jones, B.F., Wuchty, S., Uzzi, B.:
\newblock Multi-university research teams: shifting impact, geography, and
  stratification in science.
\newblock Science (322) (2008)  1259--1262

\bibitem{mika06}
Mika, P., Elfring, T., Groenewegen, P.L.M.:
\newblock Application of semantic technology for social network analysis in the
  sciences.
\newblock Scientometrics \textbf{68}(1) (2006)  3--27

\bibitem{beaverrosen1979}
deB. Beaver, D., Rosen, R.:
\newblock Studies in scientific collaboration. {Part III. P}rofessionalization
  and natural history of modem scientific coauthorship.
\newblock Scientometrics \textbf{1}(3) (1979)  231--245

\bibitem{yahoo}
{Yahoo! PlaceFinder}.
\newblock \texttt{http://developer.yahoo.com/geo/placefinder/}

\bibitem{tr-corr}
Birkholz, J.M., Bakhshi, R., Harige, R., van Steen, M., Groenewegen, P.:
\newblock Scalable analysis for large social networks: the data-aware
  mean-field approach.
\newblock Technical Report arXiv:1209.6615, CoRR (2012)

\end{thebibliography}

{}

\comment{\appendix
\section{Additional Tables}
\begin{table}[!hp]
    \setlength{\arrayrulewidth}{0.9pt}\renewcommand{\tabcolsep}{0.2cm}
  \begin{minipage}[t]{0.5\linewidth}
    \centering
    \scalebox{0.75}{
\begin{tabular}{c|c|l}
\textbf{class} & \textbf{sci. age} & \textbf{avg. \# papers} \\
\hline
NL0004 & 2010s & 5.387970137876415\\ 
NL0004 & 2000s & 5.42581655220262\\ 
NL0004 & 1990s & 5.531797518480899\\ 
NL0004 & 1980s & 5.477090962989656\\ 
NL0004 & 1970s & 5.528338394327215\\ 
NL0019 & 2010s & 5.475312984569005\\ 
NL0019 & 2000s & 5.758196664185871\\ 
NL0019 & 1990s & 6.023724882790455\\ 
NL0019 & 1980s & 6.8663700815459405\\ 
NL0019 & 1970s & 5.984149638007387\\ 
NL0013 & 2010s & 5.413644990583579\\ 
NL0013 & 2000s & 5.474109082220787\\ 
NL0013 & 1990s & 5.621564491339345\\ 
NL0013 & 1980s & 5.56704430784838\\ 
NL0013 & 1970s & 5.631497034501541\\ 
NL0032 & 2000s & 2.3746979684703073\\ 
NL0032 & 1990s & 2.3953334162184707\\ 
NL0050 & 2010s & 1.7531110121700983\\ 
NL0050 & 2000s & 1.9902354176212567\\ 
NL0050 & 1990s & 3.1478709832331986\\ 
NL0050 & 1980s & 2.446389301965944\\ 
NL0026 & 2010s & 5.463616325979629\\ 
NL0026 & 2000s & 5.891423256338538\\ 
NL0026 & 1990s & 6.3905122793287426\\ 
NL0026 & 1980s & 8.600700339873251\\ 
NL0026 & 1970s & 5.444605947535513\\ 
NL0061 & 2000s & 2.3171354951464065\\ 
NL0061 & 1990s & 1.9297842271767376\\ 
NL0014 & 2010s & 4.225951864691698\\ 
NL0014 & 2000s & 3.53224949333238\\ 
NL0014 & 1990s & 4.874423147075234\\ 
NL0014 & 1980s & 2.547490074211703\\ 
NL0014 & 1970s & 2.5172215234219197\\ 
NL0020 & 2010s & 5.638230468869671\\ 
NL0020 & 2000s & 5.739481550233624\\ 
NL0020 & 1990s & 6.078863712640379\\ 
NL0020 & 1980s & 6.604347286199442\\ 
NL0020 & 1970s & 6.672447983873101\\ 
\end{tabular}}
\end{minipage}
  \begin{minipage}[t]{0.5\linewidth}
    \centering
    \scalebox{0.75}{
\begin{tabular}{c|c|l}
\textbf{class} & \textbf{sci. age} & \textbf{avg. \# papers} \\
\hline
NL0028 & 2000s & 2.1992652044875065\\ 
NL0028 & 1990s & 3.7611830958900354\\ 
NL0028 & 1980s & 5.887778929996966\\ 
NL0028 & 1970s & 1.916486429360858\\ 
NL0068 & 2000s & 3.8240436570417975\\ 
NL0068 & 1990s & 3.8880963457664737\\ 
NL0068 & 1980s & 11.046149565013765\\ 
NL0009 & 2010s & 0.9745193168988106\\ 
NL0009 & 2000s & 1.9822319970372344\\ 
NL0009 & 1990s & 3.583446273423683\\ 
NL0009 & 1980s & 3.5567522739468425\\ 
NL0009 & 1970s & 5.18923352425846\\ 
NL0035 & 2000s & 1.9853823673436444\\ 
NL0035 & 1990s & 2.649920306864022\\ 
NL0086 & 2000s & 1.2029281222985366\\ 
NL0086 & 1990s & 2.1021646153185203\\ 
NL0030 & 2010s & 4.839688780221957\\ 
NL0030 & 2000s & 4.8193549962935744\\ 
NL0030 & 1990s & 6.3334063984522\\ 
NL0030 & 1980s & 5.248208479768112\\ 
NL0030 & 1970s & 4.773586142320594\\ 
NL0016 & 2010s & 2.49393738996642\\ 
NL0016 & 2000s & 3.7673337179394926\\ 
NL0016 & 1990s & 4.886530815370229\\ 
NL0016 & 1980s & 5.116337010797732\\ 
NL0016 & 1970s & 3.6766692089901465\\ 
NL0006 & 2000s & 0.7347930885291606\\ 
NL0006 & 1990s & 1.1441892520139676\\ 
NL0021 & 2010s & 5.882071678870509\\ 
NL0021 & 2000s & 6.023222506750733\\ 
NL0021 & 1990s & 6.6374898109359615\\ 
NL0021 & 1980s & 6.884416912837332\\ 
NL0021 & 1970s & 7.578790125859185\\ 
NL0025 & 2010s & 1.8698716348980597\\ 
NL0025 & 2000s & 2.284845226862196\\ 
NL0025 & 1990s & 2.956297121398708\\ 
NL0025 & 1980s & 3.1944309000755786\\ 
\end{tabular}}
\end{minipage}
\vspace{0.2cm}
\caption{Average papers per scientific age}
\label{tab:age-uni}
\end{table}

\begin{table}[hp!]
  \setlength{\arrayrulewidth}{0.9pt}\renewcommand{\tabcolsep}{0.2cm}
  \begin{minipage}[t]{0.5\linewidth}
    \centering
    \scalebox{0.75}{
\begin{tabular}{c|c|l}
\textbf{class $A$} & \textbf{class $B$} & \textbf{avg. \# papers between} \\
\hline
NL0004  & NL0019  & 2.263407259658148\\
NL0004  & NL0013 & 0.4944409348985507\\
NL0004  & NL0050 & 0.6580256046460706\\
NL0004 &  NL0026 & 0.37047393027047626\\
NL0004 &  NL0061 & 4.592857172934723\\
NL0004 &  NL0014 & 0.30798423769098177\\
NL0004 &  NL0020 & 2.398714227857339\\
NL0004 &  NL0028 & 0.4164640889428796\\
NL0004 &  NL0068 & 0.6061747061610265\\
NL0004 &  NL0009 & 0.6633554359577802\\
NL0004 &  NL0035 &  2.6838225144495502\\
NL0004 &  NL0030 &  2.3786523246885967\\
NL0004 &  NL0016  & 4.325481586579064\\
NL0004 &  NL0021 &   1.049071904544946\\
NL0004 &  NL0025 & 1.5680876945593265\\
NL0019 &  NL0013 & 1.263164247376606\\
NL0019 &  NL0032 & 1.4011820111760283\\
NL0019 &  NL0050 & 1.3827770823889554\\
NL0019 &  NL0026 & 1.973413511503678\\
NL0019 &  NL0014 & 0.22245162967310309\\
NL0019 &  NL0020  & 4.026561202730066\\
NL0019 &  NL0068 &  2.7450113205224453\\
NL0019 &  NL0009 &  2.2099848783755123\\
NL0019 &  NL0035 & 0.2673594397262916\\
NL0019 &  NL0030 & 0.11172136840556372\\
NL0019 &  NL0016 &  2.933493270016237\\
NL0019 &  NL0006  & 4.837153260029504\\
NL0019 &  NL0021 & 1.0036034944861343\\
NL0019 &  NL0025 & 1.5857662759274087\\
NL0013 &  NL0032  & 3.5540816217586015\\
NL0013 &  NL0050  & 5.411772797511066\\
NL0013 &  NL0026 & 1.2531918484487523\\
NL0013 &  NL0061  & 5.2454509657319255\\
NL0013 &  NL0014 & 0.07632021680371706\\
NL0013 &  NL0020 & 0.8782378750009818\\
NL0013 &  NL0028 & 1.7694984665097009\\
NL0013 &  NL0068 & 1.3772086810334625\\
NL0013 &  NL0009 & 0.6301882075378858\\
NL0013 &  NL0035 & 0.5055898105802711\\
NL0013 &  NL0086 & 0.5642383479094657\\
NL0013 &  NL0030 & 0.8378555376727728\\
NL0013 &  NL0016 & 0.7356057593307926\\
NL0013 &  NL0006 & 0.8548839485370945\\
NL0013 &  NL0021 & 0.7722536065314932\\
NL0013 &  NL0025 & 0.3257699911140513\\
NL0032 &  NL0026  & 5.690289048887382\\
NL0032 &  NL0020 & 1.2001233218988976\\
NL0032 &  NL0028 & 0.9288067821483903\\
NL0032 &  NL0009 & 0.02698279194374736\\
NL0032 &  NL0035 & 0.3763169179155553\\
NL0032 &  NL0086 & 0.2118886591850918\\
NL0032 &  NL0021 & 0.16702012542785669\\
NL0032 &  NL0025 & 0.036277940387929405\\
NL0050 &  NL0020 & 1.936954005954278\\
NL0050 &  NL0028 & 0.02072773164054517\\
NL0050 &  NL0035 & 0.03541644445312909
\end{tabular}}
  \end{minipage}
  \begin{minipage}[t]{0.5\linewidth}
\centering    
  \scalebox{0.75}{
\begin{tabular}{c|c|l}
\textbf{class $A$} & \textbf{class $B$} & \textbf{avg. \# papers between} \\
\hline
NL0050 &  NL0016 & 0.04256541335823162\\
NL0050 &  NL0021 & 1.1533007053158089\\
NL0050 &  NL0025 & 0.08509532089513992\\
NL0026 &  NL0014  & 4.8561561758914005\\
NL0026 &  NL0020 & 1.4329720180518557\\
NL0026 &  NL0028 & 0.8798112260545208\\
NL0026 &  NL0009  & 4.810684760273709\\
NL0026 &  NL0035  & 4.411840411657999\\
NL0026 &  NL0030 & 0.18459424275103944\\
NL0026 &  NL0016 & 1.7136146252333302\\
NL0026 &  NL0006  & 4.313522858056843\\
NL0026 &  NL0021 & 0.46632418474997933\\
NL0026 &  NL0025 & 1.4487786326784822\\
NL0061 &  NL0020 &  2.921188836589889\\
NL0061 &  NL0028 & 0.09561269073779911\\
NL0061 &  NL0021  & 4.772077983216558\\
NL0014 &  NL0020 & 1.824921240058736\\
NL0014 &  NL0028 & 0.11588914440358757\\
NL0014 &  NL0009 & 0.03395960031874417\\
NL0014 &  NL0035 & 0.20613212283148777\\
NL0014 &  NL0030 & 1.245053309760563\\
NL0014 &  NL0021 & 0.15648093247481415\\ 
NL0014 &  NL0025 & 0.09403039170462742\\
NL0020 &  NL0028 & 0.1811394917350658\\
NL0020 &  NL0009 & 1.1500793851576996\\
NL0020 &  NL0035  & 5.05334128327187\\
NL0020 &  NL0086  & 3.5145789273766326\\
NL0020 &  NL0030 & 1.3437113730742232\\
NL0020 &  NL0016 & 1.2399636762572337\\
NL0020 &  NL0006  & 5.523622926401512\\
NL0020 &  NL0021 & 1.2026231545075297\\
NL0028 &  NL0009 & 0.036176994375173285\\
NL0028 &  NL0035 & 0.03504757402471837\\
NL0028 &  NL0016 & 0.06389037493502961\\
NL0028 &  NL0006 & 0.0823774901330473\\
NL0028 &  NL0021 & 0.31348968936134985\\
NL0028 &  NL0025 & 0.004698085703887157\\
NL0068 &  NL0009 & 0.06766762951083505\\
NL0068 &  NL0021 & 0.5779882687355763\\
NL0068 &  NL0025 & 0.25566330393809494\\
NL0009 &  NL0035 & 0.08388778211855037\\
NL0009 &  NL0086 & 0.01110923956298006\\
NL0009 &  NL0006 & 0.009835235579214842 \\
NL0009 &  NL0021 & 0.7310525345055223\\
NL0009 &  NL0025 & 0.013179244555458474\\
NL0035 &  NL0021 &  2.532724729506115\\
NL0035 &  NL0025 & 0.01383187109381942\\
NL0086 &  NL0030 & 0.5642684985452333\\
NL0086 &  NL0016 & 0.38769682014623896\\
NL0086 &  NL0021  & 4.976244164622265\\
NL0030 &  NL0016 & 1.0501730123811992\\
NL0030 &  NL0021 & 0.5051830348930917\\
NL0030 &  NL0025 & 0.22683573470800542\\
NL0016 &  NL0021 & 1.7537386698710558\\
NL0016 &  NL0025 & 0.13959766548258615\\
NL0006 &  NL0021  & 3.067102190843901\\
NL0021 &  NL0025 & 0.4936781178361756
\end{tabular}}
\end{minipage}
\caption{Communication between Dutch institutions}
\label{sec:relations}
\end{table}}

\end{document}